\documentclass{llncs}
\usepackage{amssymb}
\usepackage{epsfig}
\usepackage{makeidx}  

\begin{document}

\begin{frontmatter}

\pagestyle{headings}  

\mainmatter              

\title{Heisenberg uncertainty relation and \\ statistical measures in the square well}

\titlerunning{Statistical measures and Heisenberg uncertainty relation}  

\author{R. L\'{o}pez-Ruiz\inst{1,3} and J. Sa\~{n}udo\inst{2,3}}

\authorrunning{L\'{o}pez-Ruiz and Sa\~{n}udo}   

\institute{Department of Computer Science, Faculty of Science, \\ 
Universidad de Zaragoza, 50009 - Zaragoza, Spain\\
\email{rilopez@unizar.es}
\and
Department of Physics, Faculty of Science, \\
Universidad de Extremadura, 06071 - Badajoz, Spain \\
\email{jsr@unex.es}
\and
BIFI, Universidad de Zaragoza, 50018 - Zaragoza, Spain
}

\maketitle              

\begin{abstract}
A non stationary state in the one-dimensional infinite square well formed by a combination 
of the ground state and the first excited one is considered. The statistical complexity
and the Fisher-Shannon entropy in position and momentum are calculated with time for this system. 
These measures are compared with the Heisenberg uncertainty relation, $\Delta x\Delta p$.
It is observed that the extreme values of $\Delta x\Delta p$ coincide in time with
extreme values of the other two statistical magnitudes. 
\end{abstract}

\end{frontmatter}

\section{Introduction}

The behavior of the statistical complexity
in time-dependent systems has not been broadly investigated.
In a previous work \cite{ijamas2012}, we have studied the statistical 
complexity $C$ in a simplified time-dependent system $\rho(x,t)$ composed of 
two one-dimensional (variable $x$) identical densities that travel 
in opposite directions with the same velocity.
The analysis of $C$ was done for two Gaussian, 
rectangular, triangular, exponential and gamma traveling densities.
Specifically, the shape of $\rho(x,t)$ presenting
the maximum and minimum $C$ was explicitly shown for all these cases. 
In this direction, other time-dependent systems have been worked out.
For instance, in \cite{calbet2001}, a gas decaying toward the asymptotic equilibrium 
state was studied. It was found that this system goes towards equilibrium by 
approaching the {\it maximum complexity path}, which is the trajectory
in distribution space formed by the distributions with the maximal
complexity. Then, from a physical point of view, it can have some interest
to study the extremal behavior of statistical magnitudes in time dependent systems.

An important statistical magnitude in quantum mechanics is the Heisenberg uncertainty 
relation $\Delta x\Delta p$, which quantifies the product of the spread $\Delta$ in the 
two conjugate variables, the space $x$ and the momentum $p$, for a wave function.
This magnitude presents some similarity with the statistical complexity and Fisher-Shannon
information that are also calculated as the product of two statistical quantities, 
one of them representing the information content of the system, and 
the other one giving an idea of how far the system is from the equilibrium. 

In this work, we calculate these statistical quantities on a simple 
time-dependent quantum system, specifically one composed by a linear combination 
of the ground state and the first excited of the one-dimensional square well.
The extreme values of these magnitudes are identified and compared among them.
Finally the conclusions are established.


\section{The time-dependent quantum system}

Let us consider a particle in a box confined in the one-dimensional interval $[0,a]$,
that is, a particle constrained in an one-dimensional infinite square well of length $a$.
The eigenvalues of the energy for this system are given by  \cite{cohen1977}
\begin{equation}
E_n={n^2\pi^2\hbar^2 \over 2ma^2} \hspace{1cm} n=1,2,\ldots
\end{equation}
and the corresponding non-degenerate states are represented by the wave functions  
\begin{equation}
\varphi_n(x)=\sqrt{2\over a}\,\sin\left({n\pi x\over a}\right).
\end{equation}

In order to study the time variation of the statistical magnitudes
we must consider a non-stationary state. For simplicity, let us take that one formed 
at time $t=0$ by the normalized linear combination of the ground state ($n=1$) and the first 
excited one ($n=2$)
\begin{equation}
\Psi(x,t=0)={1\over \sqrt{2}}\,\left(\varphi_1(x)+\varphi_2(x)\right).
\end{equation}
Up to a global phase factor, this state evolves in time in a periodic motion
expressed in the the following manner:
\begin{equation}
\Psi(x,t)={1\over \sqrt{2}}\,\left(\varphi_1(x)+e^{-iwt}\varphi_2(x)\right),
\label{eq-psi}
\end{equation}
where the angular frequency $w$ of the oscillation is given by
\begin{equation}
w={E_2-E_1\over \hbar}={3\pi^2\hbar\over 2ma^2}.
\end{equation}
The probability density of this state in position space is:
\begin{equation}
\rho(x,t)=|\Psi(x,t)|^2={1\over 2}\,\left(\varphi_1^2(x)+\varphi_2^2(x)\right)+
\varphi_1(x)\varphi_2(x)\cos(wt),
\end{equation}
and in momentum space is:
\begin{equation}
\gamma(p,t)=|\hat\Psi(p,t)|^2,
\end{equation}
where $\hat\Psi(p,t)$ is the Fourier transform of $\Psi(x,t)$.

Let us remark that $\rho(x,t)$ presents the space-time symmetry
\begin{equation}
\rho(x,t)=\rho\left(a-x,t+{\pi\over\omega}\right).
\end{equation}
This symmetry implies that all the integral quantities calculated in the interval $[0,a]$
with functions $F(\rho)$ depending on the probability density $\rho$ display 
a period equal to ${\pi\over w}$, as it can be easily checked from the expression
\begin{equation}
\int_0^aF(\rho(x,t))\,dx = \int_0^aF\left(\rho\left(x,t+{\pi\over\omega}\right)\right)\,dx.
\end{equation}
A similar symmetry property can also be checked for $\gamma(p,t)$ in the momentum space.

Now we proceed to calculate for this system the Heisenberg uncertainty relation,
the statistical complexity and the Fisher-Shannon information. 

\begin{figure}[t]
\centering \includegraphics[width=8cm]{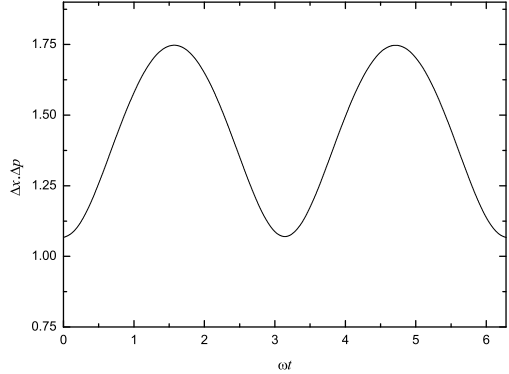}
\caption{Heisenberg uncertainty relation $\Delta x\Delta p$ versus time 
for the state $\Psi(x,t)$ considered in the text.}
\label{fig1}
\end{figure}

\section{Calculation of the statistical magnitudes}

\begin{figure}[h]
\centerline{\includegraphics[angle=0, width=6.5cm]{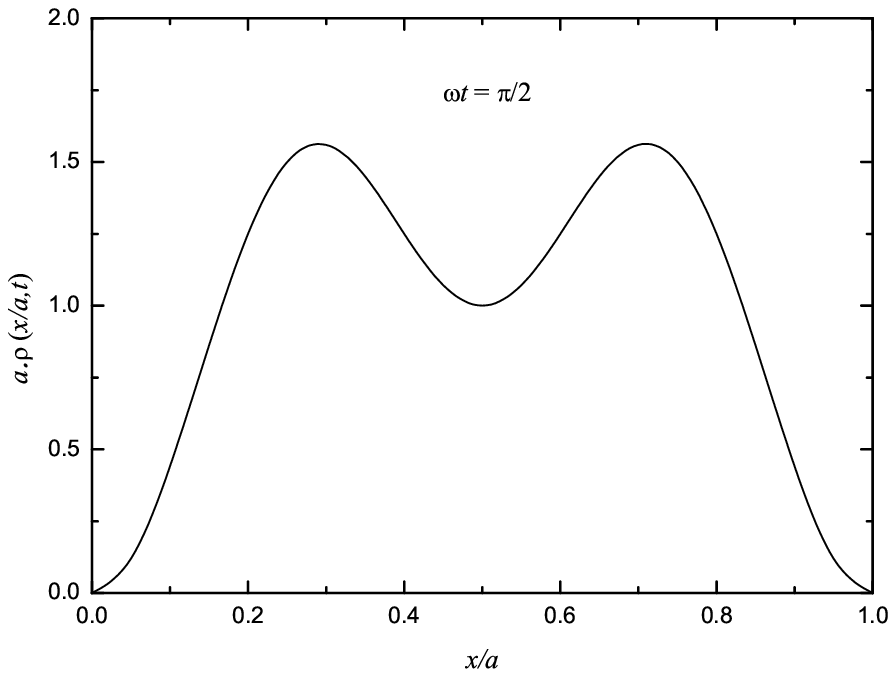}\hskip 0mm 
\includegraphics[angle=0, width=6.5cm]{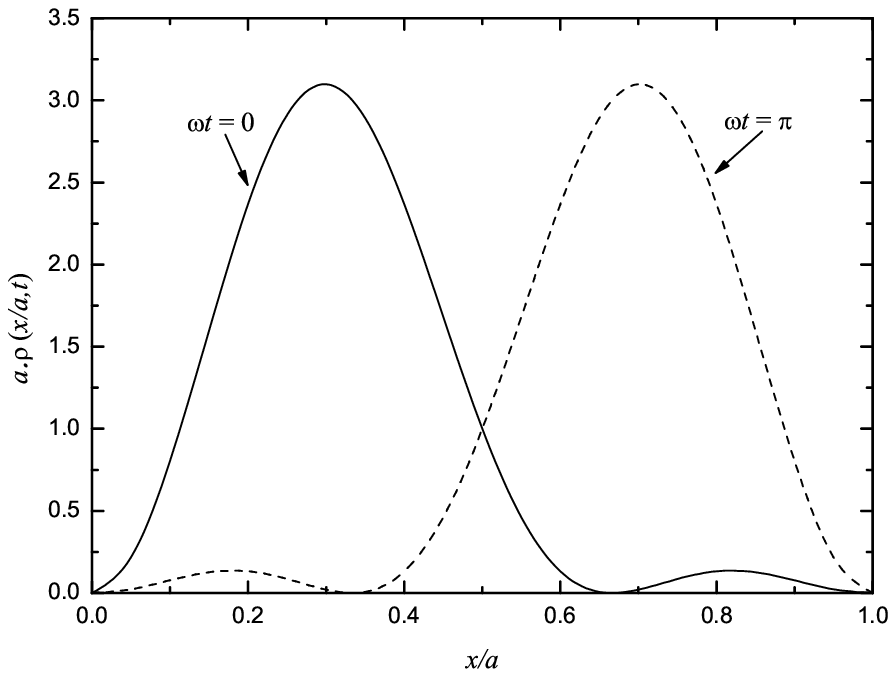}}
\centerline{(a)\hskip 6cm (b)}
\caption{Plot of $\rho(x,t)$ in adimensional units with extreme values in the uncertainty 
relation $\Delta x\Delta p$: (a) the maximum is taken at $t={\pi\over 2w}$, and 
(b) the minimum is taken at $t=0$ and $t={\pi\over w}$.}
\label{fig2}
\end{figure}

The Heisenberg uncertainty relation $\Delta x\Delta p$ is found by computing
the quantities
\begin{eqnarray}
\Delta x & = & \left(<x^2>-<x>^2\right)^{1\over 2}, \\
\Delta p & = & \left(<p^2>-<p>^2\right)^{1\over 2},
\end{eqnarray}
where $<f>$ means the average value of $f$ for the specific wave function
we are considering. In the case of the state given by Eq. (\ref{eq-psi}), 
$\Delta p$ is constant in time, $\Delta p={\hbar\pi\over a}\sqrt{3\over 2}$.
The result for the uncertainty relation,
\begin{equation}
\Delta x\Delta p ={\hbar\over 2}\left({\pi^2\over 2}-{15\over 8}-6\,
\left({16\over 9\pi}\right)^2\cos^2(wt)\right)^{1\over 2},
\end{equation}
is plotted in Fig. \ref{fig1}. Observe that  $\Delta x\Delta p$ presents 
a periodicity with period $t={\pi\over w}$, and that it shows two extreme values,
a maximum and a minimum, taken at $t={\pi\over 2w}$ and $t=0$ or ${\pi\over w}$, respectively.
The probability density $\rho$ for these extreme values is represented in Fig. \ref{fig2}.

\begin{figure}[h]
\centerline{\includegraphics[angle=0, width=6.5cm]{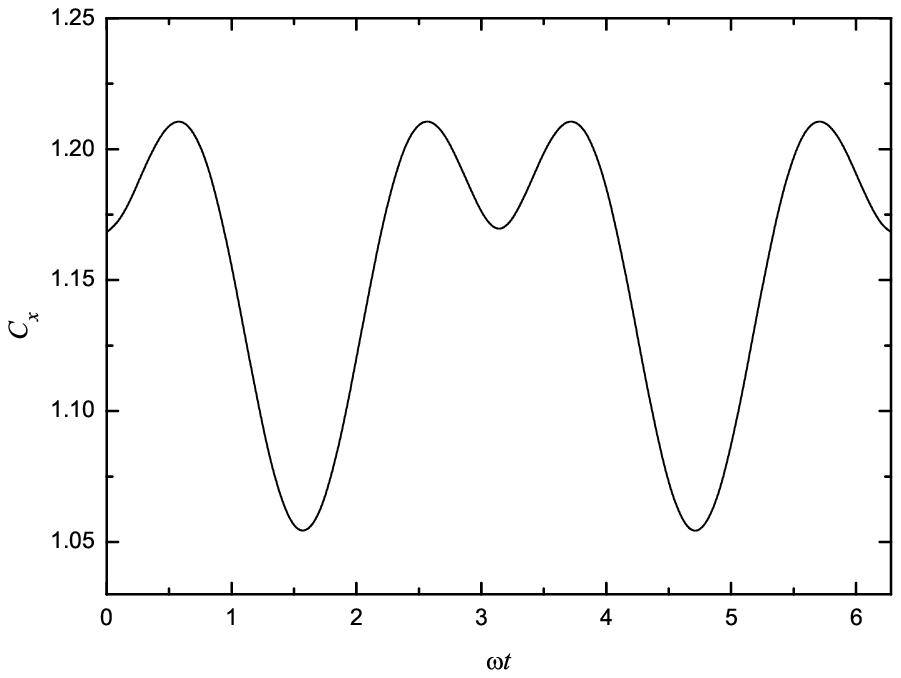} \hskip 0mm 
\includegraphics[angle=0, width=6.5cm]{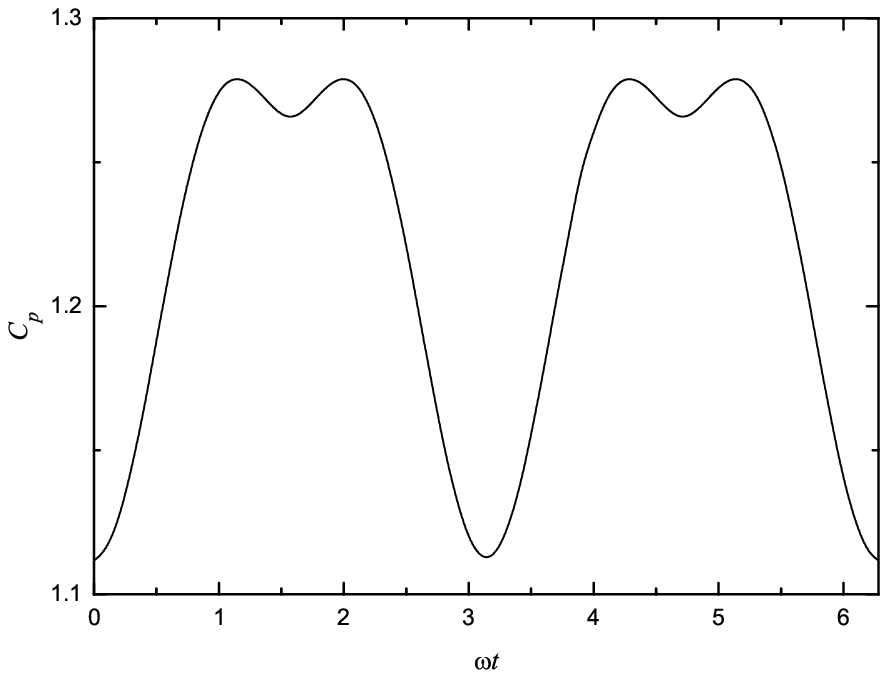}}
\centerline{(a)\hskip 6cm (b)}
\caption{Statistical complexity versus time 
for the state $\Psi(x,t)$ considered in the text: 
(a) in position, $C_x$, and (b) in momentum, $C_p$.}
\label{fig3}
\end{figure}

The statistical complexity $C$ \cite{lopez1995},
the so-called $LMC$ complexity, is defined as
\begin{equation}
C = H\cdot D\;,
\end{equation}
where $H$ represents the information content of the system and $D$ gives an idea
of how much concentrated is its spatial distribution. 
As quantifier of $H$ we take the simple exponential Shannon entropy \cite{dembo1991,lopez2002}, 
that in the position and momentum spaces takes the form, respectively,
\begin{equation}
H_x = e^{S_x}\;, \hspace{1cm}
H_p = e^{S_p}\;,
\end{equation}
where $S_x$ and $S_p$ are the Shannon information entropies \cite{shannon1948},
\begin{equation}
S_x = -\int\rho(x,t)\;\log \rho(x,t)\; dx\;, \hspace{1cm}
S_p = -\int\gamma(p,t)\;\log \gamma(p,t)\; dp\;.
\end{equation}
The disequilibrium introduced in \cite{lopez1995,lopez2002} is given by
\begin{equation}
D_x = \int\rho^2(x,t)\; dx\;, \hspace{1cm}
D_p = \int\gamma^2(p,t)\; dp\;.
\end{equation}
Then, the final expressions for $C$ in position and momentum spaces are:  
\begin{equation}
C_x = H_x\cdot D_x\;, \hspace{1cm}
C_p = H_p\cdot D_p\;.
\end{equation}
The plots of $C_x$ and $C_p$ are shown in Fig. \ref{fig3}. Observe that both
magnitudes display relative minima at $t={\pi\over 2w}$ and $t={\pi\over w}$,
just on the points where the uncertainty relation also presents relative extrema,
although in this case they are maximum and minimum, respectively.
 
Another statistical measure that has been used in quantum systems \cite{dehesa2004,sen2008} 
is the Fisher-Shannon information $P$. This quantity, in the position 
and momentum spaces, is given respectively by   
\begin{equation}
P_x = J_x\cdot I_x\;, \hspace{1cm}
P_p = J_p\cdot I_p\;,
\end{equation}
where the first factor
\begin{equation}
J_x = {1\over 2\pi e}\;e^{2S_x/3}\;, \hspace{1cm}
J_p = {1\over 2\pi e}\;e^{2S_p/3}\;,
\end{equation}
is a version of the exponential Shannon entropy \cite{dembo1991}, 
and the second factor
\begin{equation}
I_x = \int{[\nabla_x\rho(x,t)]^2\over \rho(x,t)}\; dx\;, \hspace{1cm}
I_p = \int{[\nabla_p\gamma(p,t)]^2\over \gamma(p,t)}\; dp \;,
\end{equation}
is the so-called Fisher information measure \cite{fisher1925}, that quantifies the narrowness 
of the probability density. 

\begin{figure}[h]
\centerline{\includegraphics[angle=0, width=6.5cm]{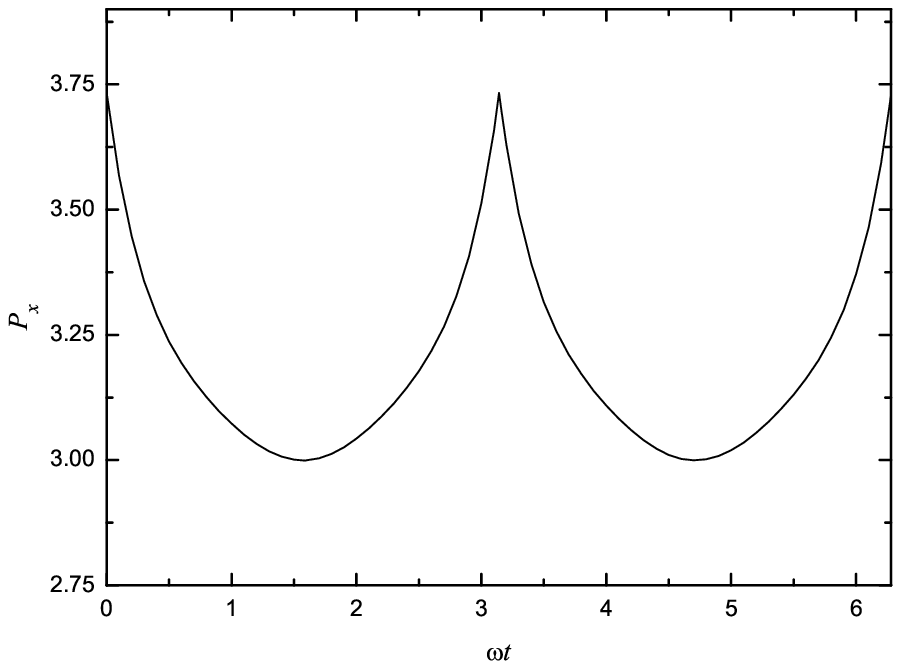} \hskip 0mm
\includegraphics[angle=0, width=6.5cm]{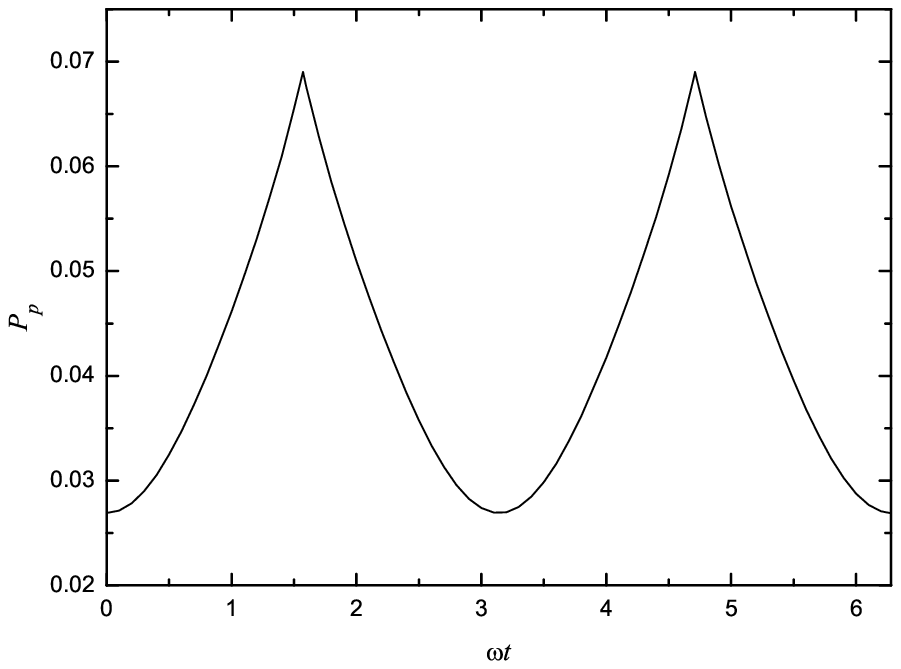}}
\centerline{(a)\hskip 6cm (b)}
\caption{Fisher-Shannon entropy versus time 
for the state $\Psi(x,t)$ considered in the text: 
(a) in position, $P_x$, and (b) in momentum, $P_p$}
\label{fig4}
\end{figure}

The plots of $P_x$ and $P_p$ are shown in Fig. \ref{fig4}. Observe that both
magnitudes display extreme values at $t={\pi\over 2w}$ and $t={\pi\over w}$,
in the same way that the uncertainty relation presents relative extrema on these points.

\section{Conclusions}
The Heisenberg uncertainty relation has been calculated for a time-dependent quantum system
in the one-dimensional square well, specifically the normalized linear combination of the ground 
state and the first excited one. This relation has a periodic behavior in time with period $\pi/w$,
and shows two extrema values that are taken at $t=\pi/(2w)$ the maximum, and at $t=0$ or $t=\pi/w$ 
the minimum. Similar properties of periodicity and extreme values are observed for the behavior of 
other statistical magnitudes, namely the statistical complexity and the Fisher-Shannon information,
that have been computed in position and momentum spaces.

\section*{Acknowledgments}
This research was supported by the spanish Grant with Ref. FIS2009-13364-C02-C01. J.S.
also thanks to the Consejer\'{\i}a de Econom\'{\i}a, Comercio e Innovaci\'on of the Junta 
de Extremadura (Spain) for financial support, Project Ref. GRU09011.

\end{document}